\documentclass[prb,twocolumn,superscriptaddress,showpacs,amsmath,amssymb,apsrev]{revtex4}

\usepackage[dvipdfmx]{graphicx}
\usepackage{dcolumn} 
\usepackage{color} 
\usepackage{dsfont}

\def\bA{{\bf A}}

\def\be{{\bf e}}

\def\bk{{\bf k}}

\def\bp{{\bf p}}
\def\bq{{\bf q}}
\def\br{{\bf r}}

\def\bQ{{\bf Q}}

\def\b0{{\bf 0}}

\def\cE{{\cal E}}

\def\cH{{\cal H}}
\def\cM{{\cal M}}

\def\cS{{\cal S}}

\def\up{\uparrow}
\def\down{\downarrow}

\def\alf{\alpha}
\def\eps{\epsilon}

\def\lam{\lambda}

\def\om{\omega}

\def\sg{\sigma}

\def\sgn{{\rm sgn}}

\begin{document}

\title{Phase stiffness in an antiferromagnetic superconductor}

\author{Walter Metzner}
\affiliation{Max Planck Institute for Solid State Research,
 D-70569 Stuttgart, Germany}
\author{Hiroyuki Yamase}
\affiliation{National Institute for Materials Science, Tsukuba 305-0047, Japan}

\date{April 12, 2019}

\begin{abstract}
We analyze the suppression of the phase stiffness in a superconductor by antiferromagnetic order. The analysis is based on a general expression for the phase stiffness in a mean-field state with coexisting spin-singlet superconductivity and spiral magnetism. N\'eel order is included as a special case. Close to half-filling, where the pairing gap is much smaller than the magnetic gap, a simple formula for the phase stiffness in terms of magnetic quasi-particle bands is derived. The phase stiffness is determined by charge carriers in small electron or hole pockets in this regime.
The general analysis is complemented by a numerical calculation for the two-dimensional Hubbard model with nearest and next-to-nearest neighbor hopping amplitudes at a moderate interaction strength. The resulting phase stiffness exhibits a striking electron-hole asymmetry. In the ground state, it is larger than the pairing gap on the hole-doped side, and smaller for electron doping. Hence, in the hole-doped regime near half-filling the ground state pairing gap sets the scale for the Kosterlitz-Thouless temperature $T_c^{\rm KT}$, while in the slightly electron-doped regime $T_c^{\rm KT}$ is determined essentially by the ground state phase stiffness. 
\end{abstract}

\maketitle


\section{Introduction}

Unconventional superconductivity in transition metal oxides and heavy fermion compounds frequently appears in the vicinity of antiferromagnetically ordered states. Antiferromagnetic fluctuations seem to provide the dominant pairing interaction in these systems. Occasionally even {\em coexistence}\/ of magnetic order and superconductivity is observed, such as in $\rm La_{2-x} Sr_x Cu O_4$, \cite{kimura99,wakimoto01} in $\rm Y Ba_2 Cu_3 O_{6+x}$, \cite{miller06,haug10} and in multilayer cuprates. \cite{mukuda12}

Coexistence of antiferromagnetism and unconventional superconductivity has also been found in various theoretical studies of the two-dimensional Hubbard model, the most popular model for the copper-oxide planes in cuprate high temperature superconductors.
At weak coupling, functional renormalization group (fRG) calculations revealed coexistence of magnetic order and $d$-wave pairing in the ground state, both in the electron and hole doped regimes. \cite{reiss07,wang14,yamase16}
Similar results were obtained from quantum cluster methods at strong coupling.\cite{lichtenstein00,capone06,aichhorn06,zheng16}
For the two-dimensional $t$-$J$ model, which is closely related to the Hubbard model in the strong coupling limit, superconductivity in coexistence with antiferromagnetic order has been found in analytic calculations based on slave boson mean-field theory \cite{yamase04} and chiral perturbation theory. \cite{sushkov04}

In two dimensional systems, the thermal phase transition between the superconducting and the normal state is of Kosterlitz-Thouless type, that is, it is associated with vortex-antivortex unbinding. The transition temperature $T_c$ is thus not necessarily determined by the size of the pairing gap, but also limited by the stiffness of the phase of the superconducting order parameter. \cite{chaikin95}
In continuum systems such as liquid Helium, the phase stiffness is proportional to a ``superfluid density'', which can be interpreted as the density of particles contributing to superfluidity. The phase stiffness in superconductors is inversely proportional to the square of the London penetration depth. The famous Uemura plot exhibits a direct proportionality between $T_c$ and $\lam_L^{-2}$ for a large number of high temperature superconductors, \cite{uemura89} indicating that $T_c$ is determined by the phase stiffness.

The undoped parent compounds of cuprate superconductors are Mott insulators. Hence, the density of charge carriers vanishes upon reducing the doping. The phase stiffness should thus be reduced, too. This yields a natural mechanism for the suppression of $T_c$ in the underdoped regime. \cite{emery95} 

Mott's metal-to-insulator transition is a strong coupling phenomenon. However, for weaker electron-electron interactions the density of charge carriers can also be reduced by antiferromagnetic order or by strong antiferromagnetic fluctuations. At half-filling, band splitting due to antiferromagnetism can turn a metal into an insulator -- known as Slater insulator. Away from but still close to half-filling, gapless charge excitations in an antiferromagnetic state are restricted to the Fermi surface of small electron or hole pockets. While the Cooper instability still exists in this situation, the phase stiffness of a superconductor is expected to be reduced by antiferromagnetism.

In this paper we present a quantitative analysis of the suppression of the phase stiffness by antiferromagnetic order. We derive a general formula for the phase stiffness in a spin-singlet superconductor coexisting with antiferromagnetism. The derivation is valid for momentum dependent gap functions of arbitrary symmetry, including $s$-wave and $d$-wave. The antiferromagnetic order is assumed to be of spiral form. This includes the N\'eel state as a special case. A formula for the phase stiffness in a superconducting ground state coexisting with N\'eel order has already been published by Tobijaszewska and Micnas. \cite{tobijaszewska05} We extend their result to spiral order with arbitrary wave vectors and to finite temperature. The formula can be substantially simplified if the pairing gap is much smaller than the magnetic gap.
Sharapov and Carbotte \cite{sharapov06} have derived a low energy theory for the influence of spin density wave order on the temperature and magnetic field dependence of the phase stiffness in a $d$-wave superconductor, where only excitations in the nodal region were taken into account. Their analysis is restricted to the special case where the magnetic wave vector connects opposite nodal points.

We evaluate the phase stiffness for the two-dimensional Hubbard model with moderate interaction strength in the weakly hole- and electron-doped regime around half-filling. To this end, we compute effective magnetic and pairing interactions from a functional renormalization group flow, and insert these into the mean-field gap equations for the ordered state. \cite{wang14,yamase16} The superconducting phase stiffness can then be calculated from the resulting magnetic and pairing order parameters. We also compute the Kosterlitz-Thouless transition temperature $T_c^{\rm KT}$. Both the phase stiffness and $T_c^{\rm KT}$ vanish upon approaching half-filling, as expected by the above qualitative arguments.
At first sight surprisingly, the Kosterlitz-Thouless temperature is significantly smaller than the mean-field transition temperature only on the electon-doped, not on the hole-doped side. Performing an analytic evaluation of the phase stiffness near half-filling, we provide a transparent explanation for this behavior.

The article is structured as follows. In Sec.~II we derive the formula for the phase stiffness in a mean-field state with coexisting spiral magnetic order and superconductivity. The simpler case of a pure BCS superconductor is discussed in the Appendix for comparison. Analytic results in special cases are obtained in Sec.~III. Finally, in Sec.~IV, we compute the phase stiffness and the Kosterlitz-Thouless temperature for the Hubbard model close to half-filling and discuss the results. A conclusion in Sec.~V closes the presentation.


\section{General formalism}

In this section we will derive a general expression for the phase stiffness in a mean-field state with coexisting spiral magnetic order and superconductivity at finite temperatures. We will be using natural units $\hbar = k_B = c = 1$.


\subsection{Mean-field action}

We consider a single-band spin-$\frac{1}{2}$ fermion system with a bare dispersion relation $\eps_\bk$ in a mean-field state with coexisting spiral magnetic and spin-singlet superconducting order. The corresponding mean-field Hamiltonian has the form
$H = H_0 + H_{\rm M} + H_{\rm SC}$, where
$H_0 = \sum_{\bp,\sg} \eps_\bp a_{\bp\sg}^\dag a_{\bp\sg}$ is the kinetic term, and 
\begin{eqnarray}
 H_{\rm M} &=& \sum_\bp A_\bp \left(
 a_{\bp+\bQ\down}^\dag a_{\bp\up} + a_{\bp\up}^\dag a_{\bp+\bQ\down} \right) , \\
 H_{\rm SC} &=& \sum_\bp \left( 
 \Delta_\bp \, a_{-\bp\down}^\dag a_{\bp\up}^\dag +
 \Delta_\bp^* \, a_{\bp\up} a_{-\bp\down} \right) .
\end{eqnarray}
Here $a_{\bp\sg}^\dag$ and $a_{\bp\sg}$ are creation and annihilation operators for fermions with momentum $\bp$ and spin orientation $\sg$, respectively. The spiral order is characterized by a wave vector $\bQ$ and a (real) magnetic gap function $A_\bp$.
Note that we have dropped a constant from the mean-field Hamiltonian which contributes to the condensation energy, but not to the electromagnetic response and phase stiffness.

In a functional integral formalism, \cite{negele87} the creation and annihilation operators are replaced by Grassmann fields $\bar\psi_{p\sg}$ and $\psi_{p\sg}$, respectively, where the index $p = (\bp,p_0)$ contains the momentum variable $\bp$ and the fermionic Matsubara frequency $p_0$. Introducing Nambu spinors
$\Psi_p = (\psi_{p\up},\bar\psi_{-p\down},\psi_{p+Q\down},\bar\psi_{-p-Q\up})$, the action corresponding to the mean-field Hamiltonian $H$ can be written as
\begin{equation} \label{action}
 \cS = - \sum_{p_0} \sum_{\bp \in {\cal M}} \bar\Psi_p G^{-1}(p) \Psi_p +
 T^{-1} \sum_\bp \xi_\bp \, ,
\end{equation}
with $\xi_\bp = \eps_\bp - \mu$, where $T$ is the temperature and $\mu$ is the chemical potential. The field independent term is generated by the anticommutation of operators required for normal ordering before passing to the functional integral representation.
The inverse propagator in Eq.~(\ref{action}) is a 4x4 matrix of the form
\begin{equation} \label{G}
G^{-1}(p) = ip_0 I - H_\bp \, , 
\end{equation}
where $I$ is the 4x4 unit matrix, and
\begin{equation} \label{H_p}
 H_\bp = \left( \begin{array}{cccc}
 \xi_\bp & -\Delta_\bp & A_\bp & 0 \\
 -\Delta_\bp^* & -\xi_{-\bp} & 0 & -A_{-\bp-\bQ} \\
 A_\bp & 0 & \xi_{\bp+\bQ} & \Delta_{-\bp-\bQ} \\
 0 & -A_{-\bp-\bQ} & \Delta_{-\bp-\bQ}^* & -\xi_{-\bp-\bQ}
 \end{array} \right) \, .
\end{equation}
The momentum summation is restricted to a reduced (magnetic) Brillouin zone ${\cal M}$, which must be chosen such that there is no double counting. For example, for spiral states with wave vectors of the form $\bQ = (\pi-2\pi\eta,\pi)$, a convenient choice of $\cal M$ is defined by restricting $p_y$ to $|p_y| \leq \pi/2$. For the N\'eel state, the magnetic zone defined by $\cos p_x + \cos p_y \geq 0$ is the most common choice.


\subsection{Coupling to electromagnetic field}

The phase stiffness of a superconductor can be computed from the linear response function which determines the current induced by an external electromagnetic field. Describing the latter by a vector potential $\bA$ in a gauge where the scalar potential $\phi$ vanishes and ${\rm div} \bA = 0$, the induced electric current density is given by
\begin{equation}
 j_\alf(\bq,\om) = - \sum_{\alf'} K_{\alf\alf'}(\bq,\om) \, A_{\alf'}(\bq,\om) \, .
\end{equation}
This defines the response function $K_{\alf\alf'}(\bq,\om)$.
The phase stiffness is related to its static limit
\begin{equation}
 K_{\alf\alf'} = \lim_{\bq \to \b0} K_{\alf\alf'}(\bq,0) \, .
\end{equation}
$K_{\alf\alf'}$ is usually diagonal, $K_{\alf\alf'} = K_\alf \delta_{\alf\alf'}$,
and the phase stiffness $J_\alf$ is given by
\begin{equation}
 K_\alf = (2e)^2 J_\alf \, .
\end{equation}
This relation follows directly from the form of the Ginzburg-Landau action for a superconductor coupled to an electromagnetic field. \cite{ashcroft76}
In isotropic systems or systems with cubic symmetry, $K_\alf$ and $J_\alf$ do not depend on the direction, that is, $K_\alf = K$ and $J_\alf = J$ are independent of $\alf$. 
In a superconductor with a parabolic dispersion $\eps_\bk = \frac{\bk^2}{2m}$, one can relate $K$ to a ``superfluid density'' $n_s$ via the formula $K = n_s e^2/m$. \cite{rickayzen80} For electrons in a crystal there is no such relation.

For electrons moving in a crystal lattice, the vector potential couples to the electrons via a phase factor multiplying the hopping amplitudes,
\begin{equation}
 t_{jj'}(\bA) = t_{jj'}
 \exp \left[ ie \int_{\br_j}^{\br_{j'}} \bA(\br,t) \cdot d\br \right] \, ,
\end{equation}
where $e$ is the electron charge (that is, negative), $\br_j$ is the position vector of the site $j$ in real space, and the integral is along a straight line from $\br_j$ to $\br_{j'}$. This yields a contribution to the action of the form \cite{voruganti92}
\begin{equation} \label{S_A}
 \cS_A = \sum_{p,p'} \sum_\sg \bar\psi_{p\sg} V_{pp'}[\bA] \psi_{p'\sg} \, ,
\end{equation}
where
\begin{eqnarray}
 V_{pp'}[\bA] &=& e \sum_\alf \eps_{\bp/2+\bp'/2}^\alf \, A_\alf(p-p') \nonumber \\
 &+& \frac{e^2}{2} \sum_{\alf,\alf'} \eps_{\bp/2+\bp'/2}^{\alf\alf'}
 \sum_k A_{\alf}(p-p'-k) A_{\alf'}(k) \nonumber \\
 &+& \dots \, ,
\end{eqnarray}
with $\eps_\bp^\alf = \partial\eps_\bp/\partial p_\alf$, and $\eps_\bp^{\alf\alf'} = \partial^2\eps_\bp/(\partial p_\alf \partial p_{\alf'})$.
Contributions beyond quadratic order in $\bA$ do not contribute to $K_{\alf\alf'}(\bq,\om)$.
For electrons in a continuum with a quadratic dispersion relation $\eps_\bp = \bp^2/(2m)$, the above expressions are the same as those obtained by minimal gauge invariant coupling.

In lattice models with density-density interactions, such as the Hubbard model or the extended Hubbard model, the vector potential couples only to the hopping amplitudes, not to the interaction terms. However, in an effective low-energy model, additional terms corresponding to vertex corrections generally appear even on mean-field level. Here we neglect these contributions. For a momentum dependent magnetic gap function $A_\bp$ this approximation violates gauge invariance.

The coupling term $\cS_A$ can also be transformed to Nambu representation.
The linear (in $\bA$) contribution to $\cS_A$ can be written as
\begin{equation}
 \cS_A^{(1)} = \sum_\alf \sum_{p_0,p'_0} \sum_{\bp,\bp' \in {\cal M}} 
 \bar\Psi_p \lam_{\bp\bp',\alf}^{(1)} \Psi_{p'} \, A_\alf(p-p') \, ,
\end{equation}
with
\begin{widetext}
\begin{equation}
 \lam_{\bp\bp',\alf}^{(1)} = e \left( \begin{array}{cccc}
 \eps_{\bp/2+\bp'/2}^\alf & 0 & 0 & \quad 0 \\ 
 0 & -\eps_{-\bp/2-\bp'/2}^\alf & 0 & \quad 0 \\
 0 & 0 & \eps_{\bp/2+\bp'/2+\bQ}^\alf & \quad 0 \\ 
 0 & 0 & 0 & -\eps_{-\bp/2-\bp'/2-\bQ}^\alf \end{array} \right) \, .
\end{equation}
A constant $e \sum_{\bp,\alf} \eps_\bp^\alf A_\alf(0)$ arising from the anticommutation of the field operators for spin-$\down$ particles vanishes due to the antisymmetry of $\eps_\bp^\alf$, and thus does not yield a contribution to $S_A^{(1)}$.
The quadratic term can be written as
\begin{eqnarray}
 \cS_A^{(2)} &=&
 \sum_{\alf,\alf'} \sum_{p_0,p'_0} \sum_{\bp,\bp' \in {\cal M}}
 \bar\Psi_p \lam_{\bp\bp',\alf\alf'}^{(2)} \Psi_{p'}
 \sum_k A_\alf(p-p'-k) A_{\alf'}(k) \nonumber \\
 &+& \frac{e^2}{2T} \sum_{\alf,\alf'} \sum_\bp \eps_\bp^{\alf\alf'}
 \sum_k A_\alf(-k) A_{\alf'}(k) \, ,
\end{eqnarray}
with
\begin{equation}
 \lam_{\bp\bp',\alf\alf'}^{(2)} = \frac{e^2}{2} \left( \begin{array}{cccc}
 \eps_{\bp/2+\bp'/2}^{\alf\alf'} & 0 & 0 & \quad 0 \\ 
 0 & -\eps_{-\bp/2-\bp'/2}^{\alf\alf'} & 0 & \quad 0 \\
 0 & 0 & \eps_{\bp/2+\bp'/2+\bQ}^{\alf\alf'} & \quad 0 \\ 
 0 & 0 & 0 & -\eps_{-\bp/2-\bp'/2-\bQ}^{\alf\alf'} \end{array} \right) \, .
\end{equation}
\end{widetext}
The second term in $S_A^{(2)}$ arises from the anticommutation of the creation and annihilation operators for spin-$\down$ particles in the Nambu representation.
Contributions where one of the original momentum variables $\bp$ and $\bp'$ lies inside $\cal M$ and the other outside $\cal M$ have been discarded. They do not contribute to $K_{\alf\alf'}(q)$ for $\bq \to 0$.


\subsection{Evaluation of response function $K$}

The current density is given by
\begin{equation}
 j_\alf(\bq,\om) =
 - \frac{1}{V} \frac{\partial\Omega[\bA]}{\partial A_\alf(-\bq,-\om)} \, ,
\end{equation}
where $\Omega[\bA]$ is the grand canonical potential in the presence of the vector potential $\bA$, and $V$ is the volume of the system.
The response function $K_{\alf\alf'}(\bq,\om)$ can thus be obtained by expanding $\Omega[\bA]$ to second order in $\bA$.
The grand canonical potential is given by the functional integral
\begin{equation}
 \Omega[\bA] = - T \ln \int D[\bar\psi,\psi] 
 e^{- \cS[\bar\psi,\psi] - \cS_A[\bar\psi,\psi,\bA]} \, .
\end{equation}

There are two distinct contributions to $\Omega[\bA]$ which are quadratic in $\bA$,
a {\em diamagnetic} contribution from the first order term in an expansion of $\Omega[\bA]$ in powers of $\cS_A$ with the second order (in $\bA$) contribution to $\cS_A$, and a {\em paramagnetic} contribution from the second order term in $\cS_A$ with the first order contribution to $\cS_A$. Both contributions are determined by a Gaussian integral.
The diamagnetic contribution is obtained as
\begin{eqnarray} \label{Om_dia}
 \Omega^{\rm dia}[\bA] &=& T \, {\rm tr} \left( G V^{(2)}[\bA] \right) \nonumber \\
 &+& \frac{e^2}{2} \sum_{\alf,\alf'} \sum_\bp \eps_\bp^{\alf\alf'}
 \sum_k A_\alf(-k) A_{\alf'}(k) \, ,
\end{eqnarray}
where
\begin{equation}
 V_{pp'}^{(2)}[\bA] = \sum_{\alf,\alf'} \lam_{\bp\bp',\alf\alf'}^{(2)}
 \sum_k A_\alf(p-p'-k) A_{\alf'}(k) \, ,
\end{equation}
and the paramagnetic contribution as
\begin{equation} \label{Om_para}
 \Omega^{\rm para}[\bA] = 
 \frac{T}{2} \, {\rm tr} \left( G V^{(1)}[\bA] G V^{(1)}[\bA] \right) \, ,
\end{equation}
where
\begin{equation}
 V_{pp'}^{(1)}[\bA] = \sum_\alf \lam_{\bp\bp',\alf}^{(1)} \, A_\alf(p-p') \, .
\end{equation}
The matrix propagator $G$ in Eqs.~(\ref{Om_dia}) and (\ref{Om_para}) is given by Eq.~(\ref{G}). The traces and matrix products involve sums over fermion momenta (restricted to $\cM$), Matsubara frequencies, and Nambu indices.

Taking derivatives with respect to $A_\alf(-q)$, one obtains the corresponding contributions to the current densities, from which one can read off the diagmagnetic and paramagnetic contributions to the response function $K_{\alf\alf'}(q)$,
\begin{eqnarray}
 K_{\alf\alf'}^{\rm dia} &=&
 \frac{2T}{V} \sum_{p_0} \sum_{\bp \in {\cal M}} {\rm tr}
 \left[ G(p) \, \lam_{\bp\bp,\alf\alf'}^{(2)} \right] \nonumber \\
 &+& e^2 \, \frac{1}{V} \sum_\bp \eps_\bp^{\alf\alf'} \, , \\
 K_{\alf\alf'}^{\rm para}(q) &=&
 \frac{T}{V} \sum_{p_0} \sum_{\bp \in {\cal M}} \nonumber \\ 
 && {\rm tr} \left[ 
 G(p) \lam_{\bp,\bp+\bq,\alf}^{(1)} G(p+q) \lam_{\bp+\bq,\bp,\alf'}^{(1)} \right] .
 \hskip 5mm
\end{eqnarray}
Here the traces and matrix products refer only to the 4x4 Nambu structure. All frequency variables are Matsubara (imaginary) frequencies so far. The diamagnetic contribution does not depend on momenta and frequencies.

The matrix propagator can be diagonalized by a unitary transformation
\begin{eqnarray}
 \tilde G(p) &=& U_\bp^\dag \, G(p) \, U_\bp \nonumber \\
 &=&  {\rm diag}\left[
 (ip_0 - E_{1\bp})^{-1},\dots,(ip_0 - E_{4\bp})^{-1} \right] , \hskip 5mm
\end{eqnarray}
where $E_{j\bp}$ are the four eigenvalues of $H_\bp$ in Eq.~(\ref{H_p}).
For real $\Delta_\bp$, the transformation matrix $U_\bp$ can be chosen real.
The traces can be evaluated with the diagonalized propagator $\tilde G(p)$ and correspondingly transformed vertices
$\tilde\lam_{\bp\bp',\alf}^{(1)} =
U_\bp^\dag \lam_{\bp\bp',\alf}^{(1)} U_{\bp'}$ and
$\tilde\lam_{\bp\bp',\alf\alf'}^{(2)} =
U_\bp^\dag \lam_{\bp\bp',\alf\alf'}^{(2)} U_{\bp'}$.
The Matsubara sums and the analytic continuation to real frequencies $\om$ can then be easily performed. Taking the limit $\bq \to 0$ after $\om \to 0$, one obtains
\begin{equation} \label{K_dia}
 K_{\alf\alf'}^{\rm dia} = 2 \int_{\bp \in {\cal M}} \sum_j
 f(E_{j\bp}) \big( \tilde\lam_{\bp\bp,\alf\alf'}^{(2)} \big)_{jj} +
 \int_{\bp} e^2 \eps_\bp^{\alf\alf'} \, ,
\end{equation}
\begin{eqnarray} \label{K_para}
 K_{\alf\alf'}^{\rm para}(\b0,0) &=&
 \int_{\bp \in {\cal M}} \sum_j f'(E_{j\bp})
 \big( \tilde\lam_{\bp\bp,\alf}^{(1)} \big)_{jj}
 \big( \tilde\lam_{\bp\bp,\alf'}^{(1)} \big)_{jj}  \nonumber \\
 &+& \int_{\bp \in {\cal M}} \sum_{j,j' \neq j}
 \frac{f(E_{j\bp}) - f(E_{j'\bp})}{E_{j\bp} - E_{j'\bp}} \nonumber \\
 && \times \big( \tilde\lam_{\bp\bp,\alf}^{(1)} \big)_{jj'}
 \big( \tilde\lam_{\bp\bp,\alf'}^{(1)} \big)_{j'j}  \, .
\end{eqnarray}
The total current response for $\bq \to 0$ and $\om \to 0$ is the sum
\begin{equation}
 K_{\alf\alf'} = K_{\alf\alf'}^{\rm dia} + K_{\alf\alf'}^{\rm para}(\b0,0) \, ,
\end{equation}
with $K_{\alf\alf'}^{\rm dia}$ from Eq.~(\ref{K_dia}) and $K_{\alf\alf'}^{\rm para}$ from Eq.~(\ref{K_para}).


\section{Analytic results} 

For N\'eel antiferromagnets the magnetic gap obeys the relation
$A_{-\bp-\bQ} = A_{\bp+\bQ} = A_\bp$. The same relation is trivially satisfied for spiral states with arbitrary $\bQ$, if $A_\bp$ is momentum independent. This is the case for mean-field solutions of models with a Hubbard interaction. \cite{igoshev10} More sophisticated fRG calculations yield magnetic gap functions with a weak momentum dependence, \cite{yamase16} such that $A_{-\bp-\bQ} = A_\bp$ is approximately valid also for non-N\'eel states.
A particularly simple formula for the phase stiffness can be derived for the ground state near half-filling, where the pairing gap much smaller than the magnetic gap.


\subsection{Diagonalization for $A_{-\bp-\bQ} = A_\bp$}

For $A_{-\bp-\bQ} = A_\bp$, the eigenvalue equation for the 4x4 matrix $H_\bp$ in Eq.~(\ref{H_p}) is biquadratic so that it can be easily solved analytically. Fixing the phase of the pairing gap such that $\Delta_\bp$ is real, one finds
$E_{1\bp} = E_\bp^+$, $E_{2\bp} = - E_\bp^+$,
$E_{3\bp} = E_\bp^-$, $E_{4\bp} = - E_\bp^-$, where
\begin{widetext}
\begin{equation} \label{E_p^pm}
 E_\bp^\pm = \sqrt{ \frac{1}{2} \left( E_\bp^2 + E_{-\bp-\bQ}^2 \right) \pm
 \frac{1}{2} \sqrt{ \left( E_\bp^2 - E_{-\bp-\bQ}^2 \right)^2 +
 4 \big[ (\xi_\bp + \xi_{-\bp-\bQ})^2 + (\Delta_\bp + \Delta_{-\bp-\bQ})^2
 \big] A_\bp^2 }} \, ,
\end{equation}
\end{widetext}
with $E_\bp = \sqrt{\xi_\bp^2 + A_\bp^2 + \Delta_\bp^2}$.
The transformation matrix can be written as
$U_\bp = (\be_{1\bp},\be_{2\bp},\be_{3\bp},\be_{4\bp})$, where $\be_{1\bp},\dots,\be_{4\bp}$ are the four normalized eigenvectors corresponding to the four eigenvalues. Their components can be chosen real.
The eigenvectors belonging to eigenvalues with opposite signs are related to each other by an exchange of the first with the second and the third with the fourth component, and a sign change in the second and fourth component. Hence, $U_\bp$ can be written in the form
\begin{equation} \label{U_p}
 U_\bp = \left( \begin{array}{cccc}
  u_\bp & v_\bp & \bar r_\bp & \bar s_\bp \\
 -v_\bp & u_\bp & - \bar s_\bp & \bar r_\bp \\
  r_\bp & s_\bp & \bar u_\bp & \bar v_\bp \\
 -s_\bp & r_\bp & - \bar v_\bp & \bar u_\bp
 \end{array} \right) \, .
\end{equation}
The normalization of the sum of squares in each line and column implies that
$u_\bp^2 + v_\bp^2 = \bar u_\bp^2 + \bar v_\bp^2$ and
$r_\bp^2 + s_\bp^2 = \bar r_\bp^2 + \bar s_\bp^2$.
The eigenvectors and hence the matrix $U_\bp$ can in principle be computed explicitly by solving the linear eigenvector equations $U_\bp \be_{j\bp} = E_{j\bp} \be_{j\bp}$ and normalizing the eigenvectors. This yields elementary but lengthy expressions.

The matrix elements of the vertices $\tilde\lam_{\bp\bp,\alf}^{(1)}$ and $\tilde\lam_{\bp\bp,\alf}^{(2)}$ depend on the matrix elements of the transformation matrix $U_\bp$. With $U_\bp$ of the form Eq.~(\ref{U_p}), the first order vertex $\tilde\lam_{\bp\bp,\alf}^{(1)} = U_\bp^\dag \lam_{\bp\bp,\alf}^{(1)} U_\bp$ is determined by only four distinct non-zero matrix elements
\begin{eqnarray}
 (\tilde\lam_{\bp\bp,\alf}^{(1)})_{11} &=&
 e \left[(u_\bp^2 + v_\bp^2) \eps_\bp^\alf +
 (r_\bp^2 + s_\bp^2) \eps_{\bp+\bQ}^\alf \right] \, , \nonumber \\
 (\tilde\lam_{\bp\bp,\alf}^{(1)})_{33} &=&
 e \left[(\bar r_\bp^2 + \bar s_\bp^2) \eps_\bp^\alf +
 (\bar u_\bp^2 + \bar v_\bp^2) \eps_{\bp+\bQ}^\alf \right] \, , \nonumber \\
 (\tilde\lam_{\bp\bp,\alf}^{(1)})_{13} &=&
 e \left[(u_\bp \bar r_\bp + v_\bp \bar s_\bp) \eps_\bp^\alf +
 (r_\bp \bar u_\bp + s_\bp \bar v_\bp) \eps_{\bp+\bQ}^\alf \right] \, , \nonumber \\
 (\tilde\lam_{\bp\bp,\alf}^{(1)})_{14} &=&
 e \left[(u_\bp \bar s_\bp - v_\bp \bar r_\bp) \eps_\bp^\alf +
 (r_\bp \bar v_\bp - s_\bp \bar u_\bp) \eps_{\bp+\bQ}^\alf \right] \, . \nonumber
\end{eqnarray}
The remaining matrix elements can be expressed in terms of these four or vanish:
\begin{eqnarray}
 (\tilde\lam_{\bp\bp,\alf}^{(1)})_{12} &=& (\tilde\lam_{\bp\bp,\alf}^{(1)})_{21} =
 (\tilde\lam_{\bp\bp,\alf}^{(1)})_{34} = (\tilde\lam_{\bp\bp,\alf}^{(1)})_{43} = 0 ,
\nonumber \\
 (\tilde\lam_{\bp\bp,\alf}^{(1)})_{22} &=& (\tilde\lam_{\bp\bp,\alf}^{(1)})_{11} ,
\nonumber \\
 (\tilde\lam_{\bp\bp,\alf}^{(1)})_{44} &=& (\tilde\lam_{\bp\bp,\alf}^{(1)})_{33} ,
\nonumber \\
 (\tilde\lam_{\bp\bp,\alf}^{(1)})_{23} &=& (\tilde\lam_{\bp\bp,\alf}^{(1)})_{32} =
-(\tilde\lam_{\bp\bp,\alf}^{(1)})_{41} = -(\tilde\lam_{\bp\bp,\alf}^{(1)})_{14} ,
\nonumber \\
 (\tilde\lam_{\bp\bp,\alf}^{(1)})_{24} &=& (\tilde\lam_{\bp\bp,\alf}^{(1)})_{42} =
 (\tilde\lam_{\bp\bp,\alf}^{(1)})_{31} = (\tilde\lam_{\bp\bp,\alf}^{(1)})_{13} .
\end{eqnarray}
From the second order vertex
$\tilde\lam_{\bp\bp,\alf\alf'}^{(2)} = U_\bp^\dag \lam_{\bp\bp,\alf\alf'}^{(2)} U_\bp$
only the diagonal elements
\begin{eqnarray}
 (\tilde\lam_{\bp\bp,\alf\alf'}^{(2)})_{11} &=&
 \frac{e^2}{2} \big[ (u_\bp^2 - v_\bp^2) \eps_\bp^{\alf\alf'} +
 (r_\bp^2 - s_\bp^2) \eps_{\bp+\bQ}^{\alf\alf'} \big] \nonumber \\
 &=& - (\tilde\lam_{\bp\bp,\alf\alf'}^{(2)})_{22} \, , \nonumber \\
 (\tilde\lam_{\bp\bp,\alf\alf'}^{(2)})_{33} &=&
 \frac{e^2}{2} \big[ (\bar r_\bp^2 - \bar s_\bp^2) \eps_\bp^{\alf\alf'} +
 (\bar u_\bp^2 - \bar v_\bp^2) \eps_{\bp+\bQ}^{\alf\alf'} \big] \nonumber \\
 &=& - (\tilde\lam_{\bp\bp,\alf\alf'}^{(2)})_{44} 
\end{eqnarray}
are needed.

Inserting the above expressions for the vertices into equations (\ref{K_dia}) and (\ref{K_para}), one obtains the current reponse function in terms of the eigenvalues $E_{j\bp}$ and the matrix elements of $U_\bp$. The formulae simplify considerably in the zero temperature limit, where $f(E_{1\bp}) = f(E_{3\bp}) = 0$, $f(E_{2\bp}) = f(E_{4\bp}) = 1$, and $f'(E_{j\bp}) = 0$ (except at special momenta associated with nodes in the pairing gap), so that
\begin{eqnarray} \label{K_diaT0}
 K_{\alf\alf'}^{\rm dia} &=& e^2 \int_\bp \eps_\bp^{\alf\alf'}
 - e^2 \int_{\bp \in \cM} \big[
 (u_\bp^2 - v_\bp^2 + \bar r_\bp^2 - \bar s_\bp^2) \eps_\bp^{\alf\alf'}
 \nonumber \\
 && + \, (r_\bp^2 - s_\bp^2 + \bar u_\bp^2 - \bar v_\bp^2) \eps_{\bp+\bQ}^{\alf\alf'}
 \big] \, ,
\end{eqnarray}
\begin{eqnarray} \label{K_paraT0}
 K_{\alf\alf'}^{\rm para} &=&
 -4 e^2 \int_{\bp \in \cM} \frac{1}{E_\bp^+ + E_\bp^-} \,
 \prod_{i=\alf,\alf'} \big[
 (u_\bp \bar s_\bp - v_\bp \bar r_\bp) \eps_\bp^i \nonumber \\
 && + \, (r_\bp \bar v_\bp - s_\bp \bar u_\bp) \eps_{\bp+\bQ}^i \big] \, .
\end{eqnarray}
Both results agree with the corresponding expressions for the phase stiffness $J_\alf = K_{\alf\alf}/(2e)^2$ in a N\'eel antiferromagnet coexisting with superconductivity reported by Tobijaszewska and Micnas \cite{tobijaszewska05} in their equations (13) and (14). The mathematical structure is the same for a spiral state with arbitrary $\bQ$.
Note that $\int_\bp \eps_\bp^{\alf\alf'} =
\int_{\bp \in \cM} \big[ \eps_\bp^{\alf\alf'} + \eps_{\bp+\bQ}^{\alf\alf'} \big]$ actually vanishes for electrons in a crystal, where $\eps_\bp$ is a periodic function.


\subsection{Phase stiffness near half-filling} \label{sec:IIIB}

Close to half-filling, the pairing gap $\Delta_\bp$ is much smaller than the magnetic gap $A_\bp$. In this situation the influence of pairing on magnetism is negligible, and we may compute the quasi-particle bands associated with the antiferromagnetic order and the transformation matrix $U_\bp$ in the limit $\Delta_\bp \to 0$. For a N\'eel state, as well as for spiral magnetic states with arbitrary momentum vectors $\bQ$, the bare band is split in two quasi-particle bands $\eps_\bp^+$ and $\eps_\bp^-$ of the form \cite{fresard91,voruganti92}
\begin{equation} \label{eps_p^pm}
 \eps_\bp^\pm = {\textstyle \frac{1}{2}} (\eps_\bp + \eps_{\bp+\bQ}) \pm
 \sqrt{\textstyle{\frac{1}{4}} (\eps_\bp - \eps_{\bp+\bQ})^2 + A_\bp^2} \, .
\end{equation}
In the following we assume $A_{-\bp-\bQ} = A_\bp$, which is always valid for a N\'eel state, and for spiral states with any wave vector if the gap function is momentum independent.

For $\Delta_\bp \to 0$, the four eigenvalues of $H_\bp$, Eq.~(\ref{H_p}), are then given by
$E_{1\bp} = E_\bp^+$, $E_{2\bp} = - E_\bp^+$,
$E_{3\bp} = E_\bp^-$, $E_{4\bp} = - E_\bp^-$, where
\begin{equation}
 E_\bp^\pm = |\xi_\bp^\pm| \quad \mbox{with} \quad \xi_\bp^\pm =
 \eps_\bp^\pm - \mu \, .
\end{equation}
In this limit, the matrix elements of the transformation matrix $U_\bp$, Eq.~(\ref{U_p}), are given by simple expressions. From the eigenvector equation for the eigenvalue $E_{1\bp} = E_\bp^+$ one obtains
\begin{eqnarray} \label{uvrs}
 u_\bp &=& \frac{A_\bp}{(\xi_\bp^+ - \xi_\bp)^2 + A_\bp^2} \,
 \Theta(\xi_\bp^+) \, , \nonumber \\
 v_\bp &=& \frac{A_\bp}{(\xi_\bp^+ - \xi_\bp)^2 + A_\bp^2} \,
 \Theta(-\xi_\bp^+) \, , \nonumber \\
 r_\bp &=& \frac{\xi_\bp^+ - \xi_\bp}{(\xi_\bp^+ - \xi_\bp)^2 + A_\bp^2} \,
 \Theta(\xi_\bp^+) \, , \nonumber \\
 s_\bp &=& \frac{\xi_\bp^+ - \xi_\bp}{(\xi_\bp^+ - \xi_\bp)^2 + A_\bp^2} \,
 \Theta(-\xi_\bp^+) \, .
\end{eqnarray}
The eigenvector equation for the eigenvalue $E_{3\bp} = E_\bp^-$ yields
\begin{eqnarray} \label{uvrs'}
 \bar u_\bp &=& \frac{\xi_\bp^- - \xi_\bp}{(\xi_\bp^- - \xi_\bp)^2 + A_\bp^2} \,
 \Theta(\xi_\bp^-) \, , \nonumber \\
 \bar v_\bp &=& \frac{\xi_\bp^- - \xi_\bp}{(\xi_\bp^- - \xi_\bp)^2 + A_\bp^2} \,
 \Theta(-\xi_\bp^-) \, , \nonumber \\
 \bar r_\bp &=& \frac{A_\bp}{(\xi_\bp^- - \xi_\bp)^2 + A_\bp^2} \,
 \Theta(\xi_\bp^-) \, , \nonumber \\
 \bar s_\bp &=& \frac{A_\bp}{(\xi_\bp^- - \xi_\bp)^2 + A_\bp^2} \, 
 \Theta(-\xi_\bp^-) \, .
\end{eqnarray}
Up to a global sign, the above expressions for the matrix elements are fixed uniquely by the eigenvector equations and the normalization of the eigenvectors.
Note that the matrix elements depend only via the difference $\eps_{\bp+\bQ} - \eps_\bp$ on the dispersion, since $\xi_\bp^\pm - \xi_\bp = h_\bp \pm (h_\bp^2 + A_\bp^2)^{1/2}$ with $h_\bp = \frac{1}{2} (\eps_{\bp+\bQ} - \eps_\bp)$. 

We will now apply the above simplified formulae for the quasi-particle bands and for the transformation matrix $U_\bp$ to evaluate the response function $K$ at and near half-filling in the ground state ($T=0$).


\subsubsection{Half-filling}

For a sufficiently large magnetic gap $A_\bp$, the lower and upper quasi-particle bands are separated by a global energy gap. At half-filling the lower band is completely filled and the upper band completely empty, that is, $\xi_\bp^- < 0$ and $\xi_\bp^+ > 0$ for all $\bp$. There is no Fermi surface and the system is an insulator.

For $\xi_\bp^- < 0$ and $\xi_\bp^+ > 0$, the matrix elements $v_\bp$, $s_\bp$, $\bar u_\bp$, and $\bar r_\bp$ vanish for $\Delta_\bp \to 0$. Inserting the expressions from Eqs.~(\ref{uvrs}) and (\ref{uvrs'}) for the remaining matrix elements into Eqs.~(\ref{K_diaT0}) and (\ref{K_paraT0}), and collecting the various terms, one obtains
\begin{eqnarray}
 K_{\alf\alf'}^{\rm dia} &=& - 2e^2 \int_{\bp \in \cM}
 \frac{h_\bp}{(h_\bp^2 + A_\bp^2)^{1/2}} \, h_\bp^{\alf\alf'} \, , \\[1mm]
 K_{\alf\alf'}^{\rm para} &=& - 2e^2 \int_{\bp \in \cM}
 \frac{A_\bp^2}{(h_\bp^2 + A_\bp^2)^{3/2}} h_\bp^\alf h_\bp^{\alf'} \, .
\end{eqnarray}
For a momentum independent $A_\bp = A$, a partial integration yields
$K_{\alf\alf'}^{\rm para} = - K_{\alf\alf'}^{\rm dia}$, that is,
$K_{\alf\alf'} = K_{\alf\alf'}^{\rm dia} + K_{\alf\alf'}^{\rm para}$ vanishes, as expected for an insulator.
For a momentum dependent magnetic gap function $A_\bp$, the diamagnetic and paramagnetic contribution do not cancel each other completely. This reflects the fact that our expressions for $K_{\alf\alf'}$ violate gauge invariance for a momentum dependent $A_\bp$.


\subsubsection{Hole doping}

We now consider the case of an electron density slightly below half-filling. The upper quasi-particle band remains empty, while the lower band is almost completely filled except for momenta near the top of the band. Hence, $\xi_\bp^+ > 0$ for all $\bp$, and $\xi_\bp^- > 0$ for momenta in small hole pockets $\cH$, while $\xi_\bp^- < 0$ for $\bp \notin \cH$.

Compared to the half-filled case, the response function $K_{\alf\alf'}$ differs only due to the opposite sign of $\xi_\bp^-$ for momenta in the hole pockets $\cH$. We thus compute the contribution from the hole pockets to $K_{\alf\alf'}$ with $\xi_\bp^- > 0$, and subtract the contribution at half-filling, where $\xi_\bp^- < 0$ for all momenta.
Inserting once again Eqs.~(\ref{uvrs}) and (\ref{uvrs'}) into Eqs.~(\ref{K_diaT0}) and (\ref{K_paraT0}), we find
\begin{eqnarray}
 \delta K_{\alf\alf'}^{\rm dia} &=&
 - e^2 \int_{\bp \in \cH} \big[
 (\bar r_\bp^2 + \bar s_\bp^2) \eps_\bp^{\alf\alf'} +
 (\bar u_\bp^2 + \bar v_\bp^2) \eps_{\bp+\bQ}^{\alf\alf'} \big]
 \nonumber \\
 &=& -2e^2 \int_{\bp \in \cH} \Big[ \eps_\bp^{\alf\alf'} +
 \frac{2(\xi_\bp^- - \xi_\bp)^2}{(\xi_\bp^- - \xi_\bp)^2 + A_\bp^2} h_\bp^{\alf\alf'}
 \Big] \hskip 7mm
\end{eqnarray}
and
\begin{eqnarray}
 \delta K_{\alf\alf'}^{\rm para} &=&
 - 4e^2 \int_{\bp \in \cH} \frac{1}{E_\bp^+ + E_\bp^-} \nonumber \\ 
 && \times \Big[
 0^2 - \prod_{i=\alf,\alf'} 
 (u_\bp \bar s_\bp \eps_\bp^i + r_\bp \bar v_\bp \eps_{\bp+\bQ}^i) \Big] 
 \nonumber \\
 &=& 2e^2 \int_{\bp \in \cH} \frac{A_\bp^2}{(h_\bp^2 + A_\bp^2)^{3/2}} \,
 h_\bp^\alf h_\bp^{\alf'} \, .
\end{eqnarray}
Using
\begin{eqnarray}
 \frac{\partial^2 \xi_\bp^-}{\partial p_\alf \partial p_{\alf'}} &=&
 \xi_\bp^{\alf\alf'} +
 \frac{2(\xi_\bp^- - \xi_\bp)^2}{(\xi_\bp^- - \xi_\bp)^2 + A_\bp^2} \,
 h_\bp^{\alf\alf'} \nonumber \\
 &-& \frac{A_\bp^2}{(h_\bp^2 + A_\bp^2)^{3/2}} \, h_\bp^\alf h_\bp^{\alf'} \, ,
\end{eqnarray}
the sum $\delta K_{\alf\alf'} =
\delta K_{\alf\alf'}^{\rm dia} + \delta K_{\alf\alf'}^{\rm para}$
can be written in the simple form
\begin{equation} \label{K_hole}
 \delta K_{\alf\alf'} = - 2e^2 \int_{\bp \in \cH}
 \frac{\partial^2 \xi_\bp^-}{\partial p_\alf \partial p_{\alf'}} \, .
\end{equation}
Since $K_{\alf\alf'}$ vanishes at half-filling, we have
$K_{\alf\alf'} = \delta K_{\alf\alf'}$ for the hole-doped system.

The result (\ref{K_hole}) has a simple interpretation.
In a BCS-superconductor the paramagnetic contribution to the response function $K_{\alf\alf'}$ vanishes at zero temperature, and the diamagnetic contribution is given by (Appendix~A)
\begin{equation} \label{K_dia_BCS}
 K_{\alf\alf'}^{\rm dia} = e^2 \int_\bp (1 - \xi_\bp/E_\bp) 
 \frac{\partial^2 \xi_\bp}{\partial p_\alf \partial p_{\alf'}} \, ,
\end{equation}
where $\xi_\bp = \eps_\bp - \mu$ and $E_\bp = (\xi_\bp^2 + \Delta_\bp^2)^{1/2}$ is the Bogoliubov quasi-particle energy in the superconductor. For a small $\Delta_\bp$ one can approximate $(1 - \xi_\bp/E_\bp) = 2 \Theta(-\xi_\bp)$, such that
\begin{equation}
 K_{\alf\alf'}^{\rm dia} = 2 e^2 \int_\bp \Theta(-\xi_\bp) \,
 \frac{\partial^2 \xi_\bp}{\partial p_\alf \partial p_{\alf'}} \, .
\end{equation}
The formula (\ref{K_hole}) thus could have been obtained from the standard BCS formula, replacing the bare dispersion $\xi_\bp$ by the dispersion of holes in the lower quasi-particle band, that is, by $-\xi_{\bp}^-$.


\subsubsection{Electron doping}

For an electron density slightly above half-filling, the lower quasi-particle band is completely filled, while the upper band is almost empty except for momenta near the bottom of the band. Hence, $\xi_\bp^- < 0$ for all $\bp$, and $\xi_\bp^+ < 0$ for momenta in small electron pockets $\cE$, while $\xi_\bp^+ > 0$ for $\bp \notin \cE$.
A calculation in complete analogy to the hole-doped case in the preceding section yields
\begin{equation} \label{K_el}
 \delta K_{\alf\alf'} = 2e^2 \int_{\bp \in \cE}
 \frac{\partial^2 \xi_\bp^+}{\partial p_\alf \partial p_{\alf'}} \, .
\end{equation}
This simple expression also agrees with the conventional BCS formula for the phase stiffness at zero temperature, for electrons moving in a quasi-particle band $\xi_\bp^+$.


\section{Results for the Hubbard model}

The formulae derived so far are do not refer to any particular microscopic model.
In this section we present results for the phase stiffness and the Kosterlitz-Thouless temperature for the two-dimensional Hubbard model. The magnetic and superconducting order parameters $A_\bp$ and $\Delta_\bp$, which enter the expressions for the response function $K_\alf$ derived in Sec.~II, are computed from mean-field equations with effective interactions as obtained from a fRG flow. \cite{wang14}
The flow is approximated by a one-loop truncation, with a frequency-independent two-particle vertex parametrized via a decomposition in charge, magnetic, and pairing channels with $s$-wave and $d$-wave form factors. \cite{husemann09}
These approximations are applicable only for sufficiently weak interactions.

We choose a dispersion relation
\begin{equation} \label{eps_p}
\eps_\bp = -2t (\cos p_x + \cos p_y) - 4t' \cos p_x \cos p_y \, , 
\end{equation}
corresponding to nearest and next-to-nearest neighbor hopping with amplitudes $t$ and $t'$ on a square lattice (with lattice constant one). In cuprate superconductors the ratio $t'/t$ is negative and ranges between $-0.15$ and $-0.35$. We will use $t$ as our unit of energy. All presented results are obtained for $t' = -0.15t$, and a Hubbard interaction $U = 3t$.


\subsection{Magnetic and pairing gap}

\begin{figure}
\centering
\includegraphics[width=8cm]{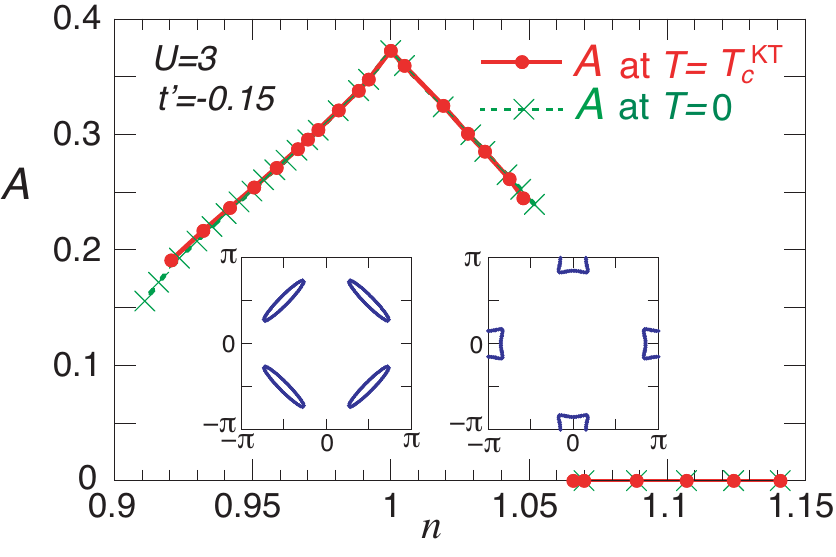} \\[3mm]
\includegraphics[width=8cm]{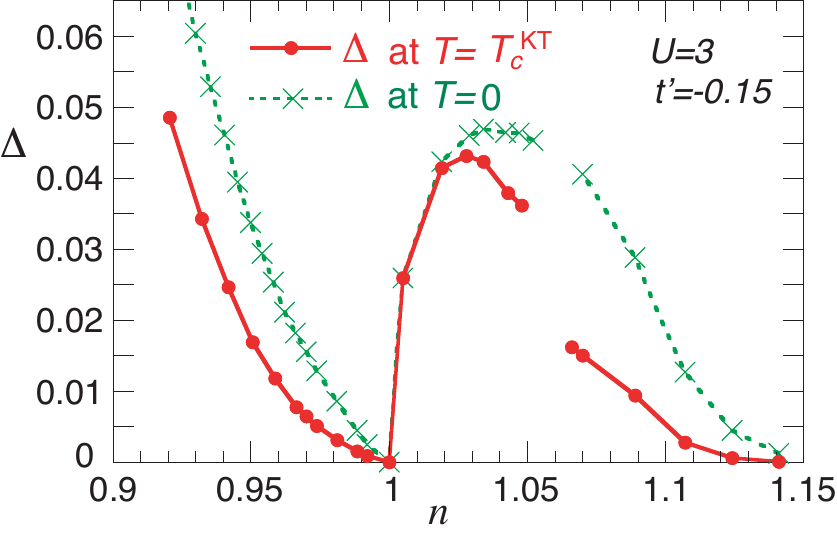}
\caption{Top: Amplitude of magnetic gap function $A = {\rm max}_\bp \, A_\bp$ as a function of electron density $n$ at $T=0$ and $T = T_c^{\rm KT}$. The inset shows hole and electron pockets for densities slightly below and slightly above half-filling ($n=1$), respectively. Bottom: Amplitude of pairing gap function $\Delta = {\rm max}_\bp \, \Delta_\bp$ as a function of electron density at $T=0$ and $T = T_c^{\rm KT}$.}
\end{figure}
In Fig.~1 we show results from the fRG calculation for the amplitudes of the magnetic gap function $A = {\rm max}_\bp \, A_\bp$ and the pairing gap function $\Delta = {\rm max}_\bp \, \Delta_\bp$, respectively, as a function of the electron density near half-filling ($n=1$). The zero temperature results have already been presented in Ref.~\onlinecite{yamase16}. The results at $T=T_c^{\rm KT}$ have been obtained by solving the mean-field equations for the gap functions at the Kosterlitz-Thouless critical temperature, which will be discussed below.
The magnetic order is of N\'eel type, with $\bQ = (\pi,\pi)$, in the density range shown. Incommensurate spiral order is found only for densities $n < 0.9$ (not shown here). \cite{yamase16} The momentum dependence of $A_\bp$ is very weak.
At half-filling the system is an insulator, while small pocket-like Fermi surfaces form in the magnetic state away from half-filling (see insets). For densities $n > 1.06$, the magnetic order vanishes, and a single large Fermi surface is recovered.
Pairing with $d$-wave symmetry is found for all densities except at half-filling. Below half-filling the largest gap is obtained for $n \approx 0.88$ (not shown here). \cite{yamase16} Close to half-filling, pairing involves mostly electrons near the pocket Fermi surfaces, where gapless excitations trigger a Cooper instability. The pairing gap increases more steeply for $n>1$ than for $n<1$ since the electron pockets are in a momentum regime where the effective interaction leading to pairing with $d_{x^2-y^2}$ symmetry is maximal. For the same reason the pairing gap on the electron doped side is practically not affected by the sudden onset of magnetic order at $n \approx 1.06$.


\subsection{Phase stiffness}

\begin{figure}
\centering
\includegraphics[width=8cm]{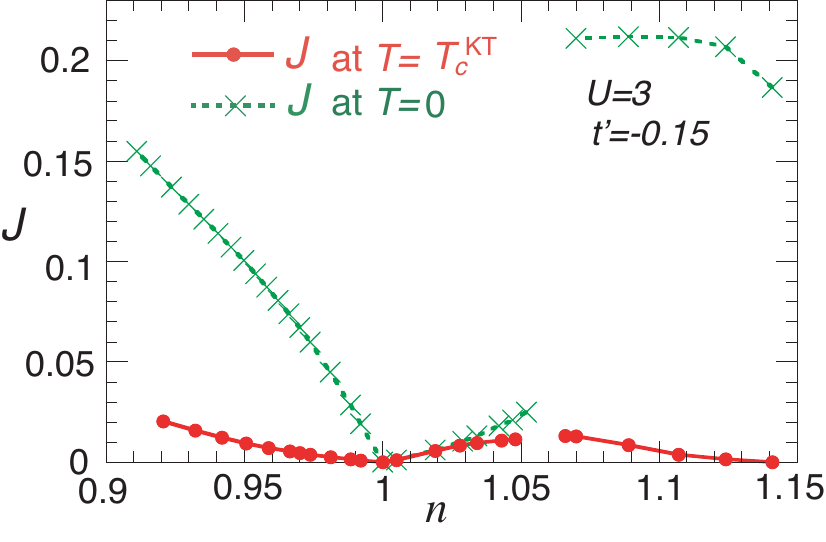}
\caption{Phase stiffness $J$ as a function of electron density at $T=0$ and $T=T_c^{KT}$.}
\end{figure}
For a $d$-wave superconductor on a square lattice, the off-diagonal elements of the tensor $K_{\alf\alf'}$ vanish, and the diagonal elements $K_{xx}$ and $K_{yy}$ are equal. This remains true also in case of coexisting N\'eel antiferromagnetic order. Hence, the phase stiffness $J = K/(2e)^2$ with $K = K_{xx} = K_{yy}$ is also just a number.
In Fig.~2 we show the phase stiffness as a function of density as obtained by evaluating the expressions (\ref{K_dia}) and (\ref{K_para}) in Sec.~II, with the gap functions computed as described above. 
Let us first discuss the ground state result ($T=0$).
The phase stiffness at $T=T_c^{\rm KT}$ is closely related to the Kosterlitz-Thouless temperature itself, as will be discussed below [see Eq.~(\ref{T_c})].
The phase stiffness vanishes at half-filling, where the pairing gap vanishes, too.
There is a striking asymmetry between the electron and hole doped side. While the pairing gap rises much more steeply upon doping on the electron-doped side, the phase stiffness increases much more steeply on the hole-doped side.
On the electron-doped side one can see that the phase stiffness jumps to a much larger value when the magnetic order disappears (discontinuously) at $n \approx 1.06$.
The drop of $J$ at $n=1.14$ is due to a tiny but finite temperature ($T=10^{-4}$) in the numerical evaluation, which suppresses $J$ if the gap is tiny, too.
Note that, in the absence of disorder, there is no critical doping at which the ground state pairing gap vanishes, since an attractive pairing interaction is present at any doping.\cite{raghu10}

Close to half-filling pairing takes place in small Fermi pockets in a robust antiferromagnetic background. The magnetic order remains practically unaffected by pairing, since $\Delta \ll A$.
In this situation one may compute the phase stiffness from the simpler formulae valid for a BCS-superconductor (see Sec.~\ref{sec:IIIB}), with the bare dispersion replaced by the magnetic quasi-particle dispersion of the partially filled band related to the Fermi pockets. 
Expanding the quasi-particle bands (\ref{eps_p^pm}) with a bare dispersion of the form (\ref{eps_p}) about their extrema, we obtain the following analytic results for the ground state phase stiffness close to half-filling
\begin{equation} \label{J_lowdop}
 J = \left\{ \begin{array}{ll} 
 (1-n) \, t^2/A \quad & \, \mbox{for} \;  n < 1 \, , \\
 (n-1) \, |t'|  \quad & \, \mbox{for} \; n > 1 \, . \end{array} \right.
\end{equation}
The derivation of these formulae is presented in Appendix B. Note that these results are valid only for $t' < 0$. For $t' > 0$ the expressions for electron and hole doping are reversed.
The derivation of Eq.~(\ref{J_lowdop}) reveals that the pronounced electron-hole asymmetry of the ground state phase stiffness in Fig.~2 is entirely due to the different curvature of the quasi-particle bands near their extrema in the electron and hole pockets.
The numerical results for the phase stiffness are consistent with Eq.~(\ref{J_lowdop}) for $n < 1 $.
For $n > 1$, the asymptotic regime near half-filling described by Eq.~(\ref{J_lowdop}) seems to be very small, so that the prefactor of the linear doping dependence of $J$ cannot be extracted from the numerical data. 


\subsection{Kosterlitz-Thouless temperature}

\begin{figure}
\centering
\includegraphics[width=8cm]{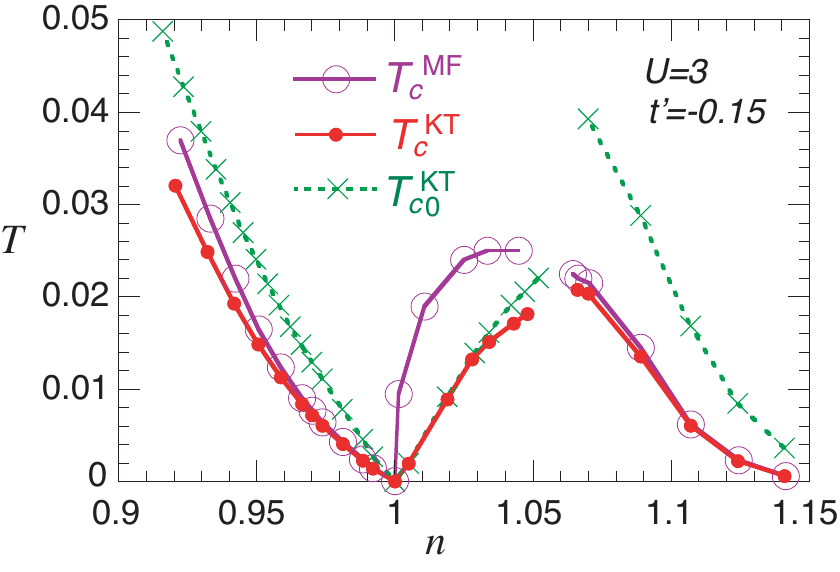}
\caption{Mean-field and Kosterlitz-Thouless critical temperatures for the superconducting phase transition as a function of electron density. At the mean-field critical temperature $T_c^{\rm MF}$ the pairing gap vanishes. The Kosterlitz-Thouless temperature $T_c^{\rm KT}$ has been computed self-consistently from the phase stiffness at $T_c^{\rm KT}$, with gap functions computed at $T_c^{\rm KT}$, while $T_{c0}^{\rm KT}$ has been obtained with ground state gaps.}
\end{figure}
In a two-dimensional system, the thermal phase transition between the superfluid and the normal phase is a Kosterlitz-Thouless transition associated with the unbinding of vortices. Magnetic order or magnetic fluctuations do not affect universal properties of this transition, as long as they are not critical at the same temperature.
The transition temperature is related to the phase stiffness by the universal relation \cite{chaikin95}
\begin{equation} \label{T_c}
 T_c^{\rm KT} = \frac{\pi}{2} J(T_c^{\rm KT}) \, ,
\end{equation}
where $J(T_c^{\rm KT})$ is the phase stiffness at the transition temperature (approached from below).

If the phase stiffness is much smaller than the pairing gap in the ground state, one may approximate $J(T_c^{\rm KT})$ in Eq.~(\ref{T_c}) by the ground state phase stiffness $J(0)$, such that $T_c^{\rm KT} \approx \frac{\pi}{2} J(0)$.
The phase stiffness generally decreases upon increasing temperature. In our mean-field theory, this decrease is partially due to the Fermi functions in Eqs.~(\ref{K_dia}) and (\ref{K_para}), and partially due to a decrease of the pairing gap.

In Fig.~3 we show a comparison of transition temperatures as obtained from three distinct approximations. $T_{c0}^{\rm KT}$ is the Kosterlitz-Thouless temperature computed from the phase stiffness $J(T_{c0}^{\rm KT})$ determined by Eqs.~(\ref{K_dia}) and (\ref{K_para}), with the ground state gaps (see Fig.~1) as input.
The other two curves are based on finite temperature solutions of the gap equations. The temperatures are still low enough so that they do not affect the fRG flow, which is stopped at the relatively high energy scale associated with the magnetic instability.
$T_c^{\rm MF}$ is the mean-field transition temperature, where the pairing gap vanishes.
$T_c^{\rm KT}$ is the Kosterlitz-Thouless temperature obtained self-consistently from the phase stiffness at $T_c^{\rm KT}$, with gap functions computed at the same temperature. The magnetic gap and the pairing gap at $T_c^{\rm KT}$ are shown in Fig.~1 together with the ground state gaps. The magnetic gap at $T_c^{\rm KT}$ is practically the same as in the ground state, since $T_c^{\rm KT}$ is much smaller than the magnetic energy scale. The phase stiffness at $T_c^{\rm KT}$ is compared to the ground state phase stiffness in Fig.~2. The temperature dependence of the phase stiffness is mostly due to the temperature dependence of the paramagnetic contribution, Eq.~(\ref{K_para}).

In the hole-doped regime, the transition temperature scale is essentially determined by the pairing gap since the ground state phase stiffness is much larger than the ground state gap. The self-consistently determined Kosterlitz-Thouless temperature $T_c^{\rm KT}$ is only slightly lower than the mean-field transition temperature. $T_{c0}^{KT}$ substantially overestimates the transition temperature, since the ground-state phase stiffness is not much smaller than the ground state gap.
The same behavior is observed in the non-magnetic regime for $n > 1.06$ on the electron doped side. By contrast, the transition temperature scale in the magnetic regime for low electron doping is limited by the phase stiffness. Here, the Kosterlitz-Thouless transition temperature computed with the ground state pairing gap is always smaller than the mean-field transition temperature at which the gap vanishes. For low electron doping, $T_c^{\rm KT}$ and $T_{c0}^{KT}$ are practically identical, and the phase stiffness at $T_c^{\rm KT}$ practically coincides with the ground state phase stiffness (see Fig.~2).


\section{Conclusion}

We have analyzed the suppression of the phase stiffness in a superconductor due to antiferromagnetic order. To this end, we have derived a general formula for the phase stiffness in a mean-field state with coexisting spin-singlet superconductivity and spiral magnetic order. Antiferromagnetic N\'eel order is included as a special case. Our formula extends an expression derived previously by Tobijaszewska and Micnas \cite{tobijaszewska05} to spiral order with arbitrary wave vectors and to finite temperatures.

We have shown analytically that close to half-filling, where the pairing gap is much smaller than the magnetic gap, the phase stiffness is given by the standard expression for a superconductor, with magnetic quasi-particle bands instead of the bare electron band. Hence, the phase stiffness near half-filling is determined by the charge carriers in small electron or hole pockets, and thus reduced.

While our general formula for the phase stiffness is valid in any dimension, it is most relevant in two dimensions, where the phase stiffness is related to the Kosterlitz-Thouless transition temperature $T_c^{\rm KT}$. We have obtained numerical results for the phase stiffness and $T_c^{\rm KT}$ in the two-dimensional Hubbard model at a moderate interaction strength ($U=3t$) around half-filling. Magnetic and pairing gap functions were computed from effective magnetic and pairing interactions as derived from an fRG flow. \cite{yamase16} The phase stiffness in the ground state exhibits a striking electron-hole asymmetry. It is much larger for hole doping than for electron doping, which can be traced back analytically to a smaller effective mass of particles in the hole-pockets, compared to the rather large effective mass in electron pockets.
On the electron-doped side, the Kosterlitz-Thouless temperature in the magnetically ordered regime is limited by the relatively small phase stiffness, compared to the ground state pairing gap. By contrast, for $U=3t$ the ground state pairing gap on the hole-doped side is smaller than the phase stiffness in the ground state, so that the Kosterlitz-Thouless temperature is only slightly lower than the mean-field transition temperature where the gap vanishes. However, this may change for stronger interactions, where the phase stiffness may decrease due to a larger magnetic (or Mott) gap, while the pairing gap is expected to increase.
A calculation with an fRG flow starting from the dynamical mean-field solution of the Hubbard model \cite{taranto14,vilardi19} could clarify the behavior at strong coupling.


\begin{acknowledgments}
We are grateful to J.~van den Brink, A.~Eberlein, J.~Mitscherling, and R.~Zeyher for valuable discussions. H.Y.\ acknowledges support by JSPS KAKENHI Grant Number JP15K05189 and JP18K18744.
\end{acknowledgments}


\begin{appendix}


\section{Phase stiffness in BCS superconductor}

Here we derive an expression for the phase stiffness of a BCS superconductor without magnetic order. This may serve as a warm-up for readers who like to follow in detail the more complicated derivation for antiferromagnetic superconductors in the main text. Derivations of the phase stiffness of superconductors presented in textbooks are usually presented for electrons in a continuum with a parabolic dispersion, which partially masks the general structure emerging for band electrons with a generic dispersion $\eps_\bk$.

In a spin-singlet BCS-superconductor, the $\bA$-independent part of the action is given by
\begin{eqnarray}
 \cS_{\rm BCS} &=& \sum_{p,\sg} (-ip_0 + \xi_\bp) \bar\psi_{p\sg} \psi_{p\sg}
 \nonumber \\
 &+& \sum_p \left( \Delta_\bp  \bar\psi_{-p\down} \bar\psi_{p\up}
 + \Delta_\bp^* \psi_{p\up} \psi_{-p\down} \right) \, ,
\end{eqnarray}
with the (generally complex) gap function $\Delta_\bp$.
Using the Nambu representation $\Psi_p = (\psi_{p\up},\bar\psi_{-p\down})$, this can be written in matrix form
\begin{equation} 
 \cS_{\rm BCS} = - \sum_p \bar\Psi_p \, G^{-1}(p) \Psi_p +
 T^{-1} \sum_\bp \xi_\bp \, ,
\end{equation}
where
\begin{equation}
 G^{-1}(p) = \left( \begin{array}{cc}
 ip_0 - \xi_\bp & \Delta_\bp \\ \Delta_\bp^* & ip_0 + \xi_{-\bp}
 \end{array} \right) \, .
\end{equation}
The field-independent term is due to the anticommutation of creation and annihiliation operators for spin-$\down$ particles in the Nambu representation (before passing to the functional integral representation). We assume $\xi_{-\bp} = \xi_\bp$.

The $\bA$-dependent part of the action $\cS_A$, Eq.~(\ref{S_A}), is also transformed to Nambu representation.
The linear (in $\bA$) term can be written as
\begin{equation}
 \cS_A^{(1)} = 
 \sum_{p,p',\alf} \bar\Psi_p \lam_{\bp\bp',\alf}^{(1)} \Psi_{p'} \, A_\alf(p-p') \, ,
\end{equation}
with
\begin{eqnarray}
 \lam_{\bp\bp',\alf}^{(1)} &=& e \left( \begin{array}{cc}
 \eps_{\bp/2+\bp'/2}^\alf & 0 \\ 0 & -\eps_{-\bp/2-\bp'/2}^\alf \end{array} \right)
 \nonumber \\
 &=& e \, \eps_{\bp/2+\bp'/2}^\alf 
 \left( \begin{array}{cc} 1 & 0 \\ 0 & 1 \end{array} \right) \, .
\end{eqnarray}
Note that a constant $e \sum_{\bp,\alf} \eps_\bp^\alf A_\alf(0)$ arising from the anticommutation of the field operators for spin-$\down$ particles vanishes due to the antisymmetry of $\eps_\bp^\alf$, and thus does not yield a contribution to $\cS_A^{(1)}$.
The quadratic term can be written as
\begin{eqnarray}
 \cS_A^{(2)} &=&
 \sum_{\alf,\alf'} \sum_{p,p'}
 \bar\Psi_p \lam_{\bp\bp',\alf\alf'}^{(2)} \Psi_{p'}
 \sum_k A_\alf(p-p'-k) A_{\alf'}(k) \nonumber \\
 &+& \frac{e^2}{2T} \sum_{\alf,\alf'} \sum_\bp \eps_\bp^{\alf\alf'}
 \sum_k A_\alf(-k) A_{\alf'}(k) \, ,
\end{eqnarray}
with
\begin{eqnarray}
 \lam_{\bp\bp',\alf\alf'}^{(2)} &=& \frac{e^2}{2} \left( \begin{array}{cc}
 \eps_{\bp/2+\bp'/2}^{\alf\alf'} & 0 \\ 0 & -\eps_{-\bp/2-\bp'/2}^{\alf\alf'}
 \end{array} \right) \nonumber \\
 &=& \frac{e^2}{2} \, \eps_{\bp/2+\bp'/2}^{\alf\alf'} 
 \left( \begin{array}{cc} 1 & 0 \\ 0 & -1 \end{array} \right) \, .
\end{eqnarray}
The second term in $\cS_A^{(2)}$ arises from the anticommutation of the creation and annihilation operators for spin-$\down$ particles in the Nambu representation.

The quadratic contributions to $\Omega[\bA]$ are then obtained as
\begin{eqnarray}
 \Omega^{\rm dia}[\bA] &=& T \, {\rm tr} \left( G V^{(2)}[\bA] \right) \nonumber \\
 &+& \frac{e^2}{2} \sum_{\alf,\alf'} \sum_\bp \eps_\bp^{\alf\alf'}
 \sum_k A_\alf(-k) A_{\alf'}(k) \, , \hskip 5mm
\end{eqnarray}
where
\begin{equation}
 V_{pp'}^{(2)}[\bA] = \sum_{\alf,\alf'} \lam_{\bp\bp',\alf\alf'}^{(2)}
 \sum_k A_\alf(p-p'-k) A_{\alf'}(k) \, ,
\end{equation}
and
\begin{equation}
 \Omega^{\rm para}[\bA] = 
 \frac{T}{2} \, {\rm tr} \left( G V^{(1)}[\bA] G V^{(1)}[\bA] \right) \, ,
\end{equation}
where
\begin{equation}
 V_{pp'}^{(1)}[\bA] = \sum_\alf \lam_{\bp\bp',\alf}^{(1)} \, A_\alf(p-p') \, .
\end{equation}
The traces sum over momentum, energy and Nambu indices.
Taking derivatives with respect to $A_\alf(-q)$, one obtains the corresponding current densities, from which one can read off the response functions
\begin{eqnarray}
 K_{\alf\alf'}^{\rm dia} &=&
 \frac{2T}{V} \sum_p {\rm tr} \left[ G(p) \, \lam_{\bp\bp,\alf\alf'}^{(2)} \right]
 + \frac{1}{V} \sum_\bp e^2 \eps_\bp^{\alf\alf'} \, , \nonumber \\ \\
 K_{\alf\alf'}^{\rm para}(q) &=&
 \frac{T}{V} \sum_p {\rm tr} \left[ 
 G(p) \lam_{\bp,\bp+\bq,\alf}^{(1)} G(p+q) \lam_{\bp+\bq,\bp,\alf'}^{(1)} \right] \, .
 \nonumber \\
\end{eqnarray}
Here the traces sum only over the Nambu indices.

The response kernel $K_{\alf\alf'}(q)$ does not depend on the global phase of $\Delta_\bp$. In the following we assume that $\Delta_\bp$ is real.
To evaluate the traces, it is convenient to choose a basis in which the propagator is diagonal in Nambu space. This is achieved by the Bogoliubov transformation
\begin{equation}
 U_\bp = 
 \left( \begin{array}{cc} u_\bp & v_\bp \\ -v_\bp & u_\bp \end{array} \right) \, ,
\end{equation}
where
\begin{equation}
 u_\bp = \frac{1}{\sqrt{2}} \sqrt{ 1 + \xi_\bp/E_\bp } \, , \;
 v_\bp = \frac{\sgn(\Delta_\bp)}{\sqrt{2}} \sqrt{ 1 - \xi_\bp/E_\bp } \, ,
\end{equation}
with the Bogoliubov energy $E_\bp = \sqrt{\xi_\bp^2 + \Delta_\bp^2}$.
The transformed Nambu propagator is given by
\begin{equation}
 \tilde G(p) = U_\bp^\dag \, G(p) \, U_\bp = \left( \begin{array}{cc} 
 (ip_0 - E_\bp)^{-1} & 0 \\ 0 & (ip_0 + E_\bp)^{-1} \end{array} \right) .
\end{equation}

The linear vertex $\lam_{\bp\bp',\alf}^{(1)}$ is proportional to the unit matrix and is therefore not affected by the Bogoliubov transformation, that is,
$\tilde\lam_{\bp\bp',\alf}^{(1)} = \lam_{\bp\bp',\alf}^{(1)}$.
The quadratic vertex $\lam_{\bp\bp',\alf\alf'}^{(2)}$ is needed only for $\bp=\bp'$, where the Bogoliubov transformation yields
\begin{equation}
 \tilde\lam_{\bp\bp,\alf\alf'}^{(2)} = \frac{e^2}{2} \eps_\bp^{\alf\alf'}
 \left( \begin{array}{cc}
 \xi_\bp/E_\bp & \Delta_\bp/E_\bp \\ \Delta_\bp/E_\bp & - \xi_\bp/E_\bp 
 \end{array} \right) \, .
\end{equation}
Since $\tilde G(p)$ is diagonal, only the diagonal elements of
$\tilde\lam_{\bp\bp,\alf\alf'}^{(2)}$ contribute to $K_{\alf\alf'}^{\rm dia}$.

Inserting $\tilde G(p)$ and $\tilde\lam_{\bp\bp,\alf\alf'}^{(2)}$ into the expression for $K_{\alf\alf'}^{\rm dia}$, and performing the Matsubara sum
$T \sum_{p_0} (ip_0 \mp E_\bp)^{-1} = f(\pm E_\bp)$, one obtains the diamagnetic contribution to $K_{\alf\alf'}$ in its final form
\begin{equation}
 K_{\alf\alf'}^{\rm dia} = e^2 \int_\bp
 \Big[ 1 - \frac{\xi_\bp}{E_\bp} + \frac{2\xi_\bp}{E_\bp} f(E_\bp) \Big] \,
 \eps_\bp^{\alf\alf'} \, .
\end{equation}
For a quadratic dispersion this reduces to
$K_{\alf\alf'}^{\rm dia} = \delta_{\alf\alf'} e^2 n/m$, where $n$ is the electron density.
Inserting $\tilde G(p)$ and $\tilde\lam_{\bp\bp,\alf}^{(1)}$ into the expression for $K_{\alf\alf'}^{\rm para}$, and performing the Matsubara sum
$T \sum_{p_0} (ip_0 \mp E_\bp)^{-1} (ip_0 \mp E_{\bp+\bq})^{-1} \to f'(E_\bp)$ for $\bq \to \b0$, one obtains the paramagnetic contribution
\begin{equation}
 K_{\alf\alf'}^{\rm para}(\b0,0) = 
 2 e^2 \int_\bp f'(E_\bp) \, \eps_\bp^\alf \eps_\bp^{\alf'} \, .
\end{equation}
The off-diagonal elements ($\alf \neq \alf'$) vanish if $\eps_\bp$ is symmetric in $p_x$ and $p_y$.
For $\Delta_\bp = 0$, a partial integration yields
$K_{\alf\alf'}^{\rm dia} + K_{\alf\alf'}^{\rm para}(\b0,0) = 0$, that is, $K_{\alf\alf'}$ vanishes.
At zero temperature one has $f(E_\bp) = 0$, so that
$K_{\alf\alf'}^{\rm dia} = e^2 \int_\bp
 \left( 1 - \xi_\bp/E_\bp \right) \, \eps_\bp^{\alf\alf'}$ and
$K_{\alf\alf'}^{\rm para}(\b0,0) = 0$.


\section{Phase stiffness near half-filling}

In this appendix we derive the analytic results Eq.~(\ref{J_lowdop}) for the phase stiffness in the two-dimensional Hubbard model close to half-filling, under the condition that the pairing gap is much smaller than the magnetic gap, such that the mean-field theory for pairing in coexistence with magnetic order can be simplified to a BCS theory for electrons moving in the quasi-particle bands $\eps_\bp^\pm$ associated with the magnetic order (see Sec.~\ref{sec:IIIB}).
We assume N\'eel order, that is, $\bQ = (\pi,\pi)$ near half-filling.


\subsection{Hole doping}

For electron densities smaller than but close to half-filling, the lower quasi-particle band $\eps_\bp^-$ is partially filled, with empty states (holes) only near the maxima of the band, while the upper quasi-particle band $\eps_\bp^+$ is completely empty.
For a bare dispersion of the form Eq.~(\ref{eps_p}) with $t > 0$ and $t' < 0$, the maxima of the lower quasi-particle band $\eps_\bp^-$ in Eq.~(\ref{eps_p^pm}) are situated at the four symmetric points
$\pm (\pi/2,\pm \pi/2)$ in the Brillouin zone, if $|t'/t|$ is not unrealistically large.
Neglecting the generally weak momentum dependence of $A_\bp$, and expanding for example around the maximum at $(\pi/2,\pi/2)$, one finds
\begin{equation}
 \eps_\bp^- = - A - \frac{2t^2}{A} (\delta p_x + \delta p_y)^2
              - 4t' \delta p_x \delta p_y \, ,
\end{equation}
where $\delta p_\alf = p_\alf - \pi/2$ for $\alf=x,y$. The matrix of second derivatives of $\xi_\bp^- = \eps_\bp^- - \mu$ is thus given by
\begin{equation}
 \left( \frac{\partial^2 \xi_\bp^-}{\partial p_\alf \partial p_{\alf'}} \right) =
 - 4 \left( \begin{array}{cc} t^2/A & t^2/A + t' \\ t^2/A + t' & t^2/A
 \end{array} \right) \, . 
\end{equation}
Both eigenvalues of this matrix, $-(8t^2/A + 4t')$ and $4t'$, are negative if $t'<0$ and
$|t'/t| < 2t/A$.

The signs of the diagonal elements of the matrix of derivatives are the same for all four extrema, while the off-diagonal elements at the extrema $(\pm \pi/2, \mp \pi/2)$ are negative.
Hence, the latter cancel when summing over all four pockets in Eq.~(\ref{K_hole}), such that $K_{\alf\alf'} = K \delta_{\alf\alf'}$, where
\begin{equation}
 K = - 2e^2 \int_{\bp \in \cH} \frac{- 4t^2}{A} = \frac{4e^2 t^2}{A} (1-n) \, .
\end{equation}
The phase stiffness $J = K/(4e^2)$ is thus obtained as $J = (1-n) t^2/A$.


\subsection{Electron doping}

For electron densities larger than but close to half-filling, the upper quasi-particle band $\eps_\bp^+$ is partially filled, with occupied states only near the bottom of the band, while the lower quasi-particle band $\eps_\bp^-$ is completely filled.
For a bare dispersion of the form Eq.~(\ref{eps_p}) with $t > 0$ and $t' < 0$, the minima of the upper quasi-particle band $\eps_\bp^+$ in Eq.~(\ref{eps_p^pm}) are situated at the two points $(\pi,0)$ and $(0,\pi)$ in the Brillouin zone.
Neglecting the momentum dependence of $A_\bp$, and expanding for example around the maximum at $(\pi,0)$, one finds
\begin{equation}
 \eps_\bp^+ = 4t' + A - 2t' (\delta p_x^2 + \delta p_y^2) \, ,
\end{equation}
where $\delta p_x = p_x - \pi$ and $\delta p_y = p_y$. The matrix of second derivatives of $\xi_\bp^+ = \eps_\bp^+ - \mu$ is thus given by
\begin{equation}
 \left( \frac{\partial^2 \xi_\bp^+}{\partial p_\alf \partial p_{\alf'}} \right) =
 - 4 \left( \begin{array}{cc} t' & 0 \\ 0 & t'
 \end{array} \right) \, . 
\end{equation}
Inserting this in Eq.~(\ref{K_el}) yields $K_{\alf\alf'} = K \delta_{\alf\alf'}$, where
\begin{equation}
 K = 2e^2 \int_{\bp \in \cE} (-4t') = 4e^2 |t'| (n-1) \, .
\end{equation}
The phase stiffness $J = K/(4e^2)$ is thus obtained as $J = (n-1) |t'|$.

\end{appendix}


\end{document}